\documentclass[11pt,a4paper]{article}
\pdfoutput=1
\usepackage{jheppub}
\usepackage{tikz}
\usetikzlibrary{intersections, calc, arrows.meta, bending}
\usepackage{simpler-wick}
\usepackage{orcidlink}
\usepackage{bbm}
\usepackage{booktabs}

\DeclareMathOperator{\Tr}{Tr}

\newcommand{\ri}{\mathrm{i}}
\renewcommand{\th}{\theta}

\newcommand{\cob}{\delta}

\newcommand{\hf}{\frac{1}{2}}

\newcommand{\si}{\sigma}
\newcommand{\Si}{\Sigma}

\newcommand{\lap}{\Delta}
\newcommand{\bra}{\langle}
\newcommand{\ket}{\rangle}
\newcommand{\la}{\lambda}

\newcommand{\h}[1]{\widehat{#1}}
\newcommand{\bt}{\beta}

\newcommand{\Om}{\Omega}
\newcommand{\rt}[1]{\sqrt{#1}}
\newcommand{\cO}{\mathcal{O}}

\newcommand{\cF}{\mathcal{F}}

\newcommand{\cD}{\mathcal{D}}

\newcommand{\cC}{\mathcal{C}}

\makeatletter
\gdef\@fpheader{}
\makeatother

\begin{document}
\title{ASEP/DSSYK duality and strange correlator}

\author{Kazumi Okuyama\,\orcidlink{0000-0002-3591-6954}\,}
\affiliation{Department of Physics, 
Shinshu University\\ 3-1-1 Asahi, Matsumoto 390-8621, Japan}

\emailAdd{kazumi@azusa.shinshu-u.ac.jp}

\abstract{
We show that the moment of the transfer matrix of 
the double scaled SYK model 
is written as an overlap between the stationary state of 
ASEP (asymmetric simple exclusion process) and a product state.
We argue that this overlap is an analogue of the strange correlator
appearing in the correspondence between the Levin-Wen string-net model and the
Turaev-Viro state sum.
}

\maketitle

\section{Introduction and summary}
The double scaled SYK (DSSYK) model \cite{Cotler:2016fpe,Berkooz:2018jqr} 
is a useful toy model for the study of
quantum gravity and holography. As shown in \cite{Berkooz:2018jqr}, 
DSSYK is exactly solvable by using the
technique of the chord diagrams and the transfer matrix.
Curiously, it is observed
in \cite{Okuyama:2023bch,Okuyama:2023byh,Okuyama:2025ebn,Watanabe:2025rwp} 
that the same transfer matrix also appears in a completely different model, known
as the asymmetric simple exclusion process (ASEP) \cite{derrida1993exact,blythe2000exact}. 
ASEP is a 1d lattice gas model of particles 
hopping asymmetrically to the left and right with hard core exclusion imposed.
The probability of a configuration of ASEP
obeys a Markov process, and its Markov matrix turns out to be a Hamiltonian of the
integrable open XXZ spin chain \cite{Crampe:2014aoa}.
The stationary state of ASEP is given by a matrix
product state (MPS) of the spin chain.
Our key observation is that the moment $Z=\bra0|T^k|0\ket$ of the transfer matrix 
$T$ of DSSYK is written as an overlap of the stationary state $|\Psi\ket$ of ASEP 
and some product state $|\Om\ket$
\begin{equation}
\begin{aligned}
 Z=\bra\Om|\Psi\ket.
\end{aligned} 
\label{eq:strange}
\end{equation}
In this paper, we will call this relation ``ASEP/DSSYK duality''.

\begin{figure}[bth]
\centering
\begin{tikzpicture}[scale=1]
\fill[yellow] (-3,0)--(3,0)--(3,2)--(-3,2)--cycle;
\draw[ultra thick] (-3,0)--(3,0);
\draw[-Latex,thick] (-1.5,-0.2)--(1.5,-0.2);
\draw (0,-0.2) node [below]{$T$};
\draw[-Latex,thick] (0,0.3)--(0,1.7);
\draw (0,1) node [right]{$H$};
\draw (3,0) node [right]{$|\Psi\ket$~physical boundary};
\draw (3,2) node [right]{$\bra\Omega|$~symmetry boundary};
\draw (2.7,1) node [left]{bulk};
\end{tikzpicture}
  \caption{
Sandwich construction of $Z=\bra\Om|\Psi\ket$
}
  \label{fig:sandwich}
\end{figure}

Interestingly, the overlap of $\bra\Om|$ and $|\Psi\ket$ in \eqref{eq:strange}
also appears in the Levin-Wen string-net model \cite{Levin:2004mi}
and this overlap is known as the strange correlator
\cite{You2013WaveFA,Vanhove:2018wlb}.
In this case, $|\Psi\ket$ is the string-net state written as a PEPS (projected entangled pair state), $|\Om\ket$ is 
some product state, and $Z$ on the left hand side of \eqref{eq:strange}
is equal to the Turaev-Viro state sum on the 3-manifold $\Si\times I$
\cite{Kadar:2009fs,Koenig:2010uua,kirillov2011stringnetmo}, where $\Si$ is a two-dimensional surface on which the string-net lives and $I=[0,1]$ is the interval.
In this paper, we will call this relation ``Turaev-Viro/string-net correspondence''.
This is reminiscent of the sandwich construction in SymTFT
(see Figure \ref{fig:sandwich}), which
has been studied extensively in the literature under the name of ``topological holography''
(see e.g. \cite{Huang:2023pyk,Vanhove:2024lmj,Zhang:2025bsm} and references therein).

The purpose of this paper is to better understand the analogy between
the ASEP/DSSYK duality and the Turaev-Viro/string-net correspondence, as
summarized in Table \ref{tab:correspondence}.
We expect
that we can draw some lessons for the holography of 3d gravity
by using this analogy with the lower dimensional holography of DSSYK.
\begin{table}[htb]
\centering
\begin{tabular}{@{}r|c|c}
\toprule
& \textbf{ASEP/DSSYK} & \textbf{Turaev-Viro/string-net}\\
 \midrule
state $|\Psi\ket$& stationary state of ASEP & Levin-Wen string-net state\\
bulk Hamiltonian $H$& XXZ spin chain Hamiltonian & Levin-Wen Hamiltonian \\
characterization of $|\Psi\ket$ & $H|\Psi\ket=0$ & $H|\Psi\ket=0$\\
tensor network for $|\Psi\ket$ & MPS & PEPS\\
partition function & $Z=\bra\Om|\Psi\ket$ & $Z=\bra\Om|\Psi\ket$\\
bulk topology & disk & $\Si\times I$ \\
sum over topologies & $T\to\text{random matrix}$ & random tensor network?\\
 \bottomrule
\end{tabular}
\caption{ASEP/DSSYK duality and Turaev-Viro/string-net correspondence}
  \label{tab:correspondence}
\end{table}

This paper is organized as follows. In section \ref{sec:ASEP},
we discuss the relation between DSSYK and ASEP and
show that the moment of the transfer matrix of DSSYK
is written as a strange correlator \eqref{eq:strange}. 
In section \ref{sec:TV}, we consider the Turaev-Viro/string-net correspondence
\eqref{eq:ZTV-strange} and point out the similarity to the ASEP/DSSYK duality. 
We conclude in section \ref{sec:discussion}
with some discussions on future problems.

\section{ASEP/DSSYK duality}\label{sec:ASEP}

\subsection{Review of DSSYK}
In this subsection, we will briefly review the definition of 
DSSYK and its solution in terms of the transfer matrix \cite{Berkooz:2018jqr}.
The Hamiltonian of the SYK model is given by
the $p$-body interaction of 
$N$ Majorana fermions $\psi_i~(i=1,\cdots,N)$
\cite{Sachdev1993,Kitaev1,Kitaev2}
\begin{equation}
\begin{aligned}
 H_{\text{SYK}}=\ri^{p/2}\sum_{1\leq i_q<\cdots< i_p\leq N}
J_{i_1\cdots i_p}\psi_{i_1}\cdots \psi_{i_p}
\end{aligned} 
\end{equation}
where 
$J_{i_1\cdots i_p}$ is a Gaussian random coupling
with zero mean and the variance
\begin{equation}
\begin{aligned}
 \bra J^2_{i_1\cdots i_p}\ket=\frac{p!(N-p)!}{N!}.
\end{aligned} 
\end{equation}
DSSYK is defined by the scaling limit
\begin{equation}
\begin{aligned}
 N,p\to\infty\quad\text{with}\quad\la=\frac{2p^2}{N}:
~\text{fixed}.
\end{aligned}
\label{eq:limit} 
\end{equation}
The expectation value of the moment $\bra \Tr (H_{\text{SYK}})^k\ket$
averaged over the random coupling $J_{i_1\cdots i_p}$ can be 
evaluated by the Wick contraction since we assumed that
the coupling $J_{i_1\cdots i_p}$ is Gaussian random.
For each Wick contraction of $J_{i_1\cdots i_p}$, we assign a chord
which connects two $H_{\text{SYK}}$'s in the trace $\Tr (H_{\text{SYK}})^k$.
In the scaling limit \eqref{eq:limit},
the remaining trace over the Hilbert space of Majorana fermions
boils down to the computation of the intersection number of chords
with the weight factor $q=e^{-\la}$.
This counting problem of intersection numbers of chord diagrams
can be solved by introducing the transfer matrix $T$
\begin{equation}
\begin{aligned}
 T=a_{+}+a_{-}
\end{aligned} 
\label{eq:T}
\end{equation}
where $a_\pm$ is the $q$-deformed oscillator obeying
the relation
\begin{equation}
\begin{aligned}
 a_{-}a_{+}-qa_{+}a_{-}=1.
\end{aligned} 
\label{eq:q-osci}
\end{equation}
Then the moment $\bra \Tr (H_{\text{SYK}})^k\ket$ is written as
\begin{equation}
\begin{aligned}
\bra \Tr (H_{\text{SYK}})^k\ket=\bra 0|T^k|0\ket,
\end{aligned} 
\label{eq:moment}
\end{equation}
where $|n\ket~(n=0,1,\cdots)$
is the so-called chord number state which represents the state with
$n$ chords
and $a_\pm$ act as the creation/annihilation operators of chords
\begin{equation}
\begin{aligned}
a_{+}|n\ket=\rt{\frac{1-q^{n+1}}{1-q}}|n+1\ket,\qquad
 a_{-}|n\ket=\rt{\frac{1-q^n}{1-q}}|n-1\ket.
\end{aligned} 
\label{eq:apm-n}
\end{equation}
The $0$-chord state is annihilated by $a_{-}$
\begin{equation}
\begin{aligned}
 a_{-}|0\ket=0,
\end{aligned} 
\end{equation}
and the inner product of chord number states is normalized as 
$\bra n|m\ket=\cob_{n,m}$.
The transfer matrix $T$ is diagonalized in the $|\th\ket$-basis
\begin{equation}
\begin{aligned}
 T|\th\ket=E(\th)|\th\ket,\quad (0\leq\th\leq\pi),
\end{aligned} 
\end{equation}
where the eigenvalue $E(\th)$ is given by
\begin{equation}
\begin{aligned}
 E(\th)=E_0\cos\th,\qquad  E_0=\frac{2}{\rt{1-q}}.
\end{aligned} 
\label{eq:Eth}
\end{equation}
The overlap of the chord number state $|n\ket$ and the eigenstate $|\th\ket$
of $T$ is given by the 
$q$-Hermite polynomial $H_n(x|q)$
with degree $n$
\begin{equation}
\begin{aligned}
\bra n|\th\ket=\bra\th|n\ket=\frac{H_n(\cos\th|q)}{\rt{(q;q)_n}}, 
\end{aligned} 
\end{equation}
which is orthogonal with respect to the measure 
$\mu(\th)=(q,e^{\pm2\ri\th};q)_\infty$ \footnote{
The $q$-Pochhammer symbol is defined by
\begin{equation}
\begin{aligned}
 (a;q)_n=\prod_{k=0}^{n-1} (1-aq^k).
\end{aligned} 
\end{equation}
We also use the notation
\begin{equation}
\begin{aligned}
 \quad
(a_1,\cdots,a_s;q)_\infty =\prod_{i=1}^s (a_i;q)_\infty,\qquad
(a^\pm;q)_\infty=(a^+;q)_\infty(a^{-};q)_\infty.
\end{aligned} 
\end{equation}}
\begin{equation}
\begin{aligned}
 \int_0^\pi\frac{d\th}{2\pi}\mu(\th)\bra n|\th\ket\bra\th|m\ket=\bra n|m\ket
=\cob_{n,m}.
\end{aligned} 
\end{equation}
The disk partition function $Z(\bt)$ of DSSYK is written as
\begin{equation}
\begin{aligned} 
Z(\bt)=\bigl\bra\Tr e^{-\bt H_{\text{SYK}}}\bigr\ket=\bra0|e^{-\bt T}|0\ket.
\end{aligned} 
\label{eq:Z-disk}
\end{equation}

We can also consider the matter operator 
$\cO_{\lap}$ with the dimension $\lap=s/p$
\begin{equation}
\begin{aligned}
 \cO_\lap=\ri^{s/2}\sum_{1\leq i_1<\cdots<i_s\leq N}K_{i_1\cdots i_s}
\psi_{i_1}\cdots\psi_{i_s}.
\end{aligned} 
\end{equation}
Assuming that $K_{i_1\cdots i_s}$ is another Gaussian random coupling independent
of $J_{i_1\cdots i_p}$, the correlator of $\cO_{\lap}$'s
can be computed by the technique of the chord diagram
as well.
In this case, there appear two types of chords, the $H$-chord and the matter chord,
coming from the Wick contractions of $J$ and $K$, respectively.
For instance, the thermal two-point function of the operator $\cO_\lap$ is written as
\begin{equation}
\begin{aligned}
 \Bigl\bra \Tr\bigl[e^{-\bt_1H}\cO_\lap e^{-\bt_2H}\cO_\lap\bigr]\Bigr\ket
=\bra 0|e^{-\bt_1T}q^{\lap\h{n}}e^{-\bt_2T}|0\ket,
\end{aligned} 
\label{eq:2pt}
\end{equation}
where $\h{n}$ is the number operator
\begin{equation}
\begin{aligned}
 \h{n}|n\ket=n|n\ket,\quad (n=0,1,\cdots).
\end{aligned} 
\label{eq:number}
\end{equation}
More generally, we can consider the bi-local operator $\cD_{\bt,\lap}$
\begin{equation}
\begin{aligned}
 \cD_{\bt,\lap}=\wick{\c \cO_\lap e^{-\bt H_{\text{SYK}}}\c \cO_\lap},
\end{aligned} 
\end{equation}
where the overline denotes
the Wick contraction of random coupling $K_{i_1\cdots i_s}$.
As shown in \cite{Berkooz:2018jqr,Okuyama:2024yya}, 
this operator is given by
\begin{equation}
\begin{aligned}
 \cD_{\bt,\lap}=\sum_{\ell=0}^\infty
(1-q)^\ell\frac{(q^{2\lap};q)_\ell}{(q;q)_\ell}
a_+^\ell  q^{\lap\h{n}}e^{-\bt T}q^{\lap\h{n}} a_-^\ell.
\end{aligned} 
\end{equation}
It turns out that this operator commutes with $T$
\begin{equation}
\begin{aligned}
 [\cD_{\bt,\lap},T]=0,
\end{aligned} 
\label{eq:bilocal-T}
\end{equation}
and hence $T$ and $\cD_{\bt,\lap}$ can be simultaneously 
diagonalized in the basis $|\th\ket$. The eigenvalue of
$\cD_{\bt,\lap}$ is given by
\begin{equation}
\begin{aligned}
 \cD_{\bt,\lap}|\th\ket=\cD_{\bt,\lap}(\th)|\th\ket,\qquad
\cD_{\bt,\lap}(\th)=\bra\th|q^{\lap\h{n}}e^{-\bt T}|0\ket.
\end{aligned} 
\end{equation}
Note that when $\bt=0$,  
$\cD_{\bt,\lap}$ reduces to the identity operator
\begin{equation}
\begin{aligned}
 \cD_{0,\lap}=\mathbbm{1}.
\end{aligned} 
\end{equation}

\subsection{ASEP and spin chain}
In \cite{Okuyama:2023bch,Okuyama:2023byh,Okuyama:2025ebn,Watanabe:2025rwp},
it was made a curious observation that
the $q$-oscillator representation of the transfer matrix 
also appears in
a statistical mechanical problem known as 
the asymmetric simple exclusion process (ASEP)
\cite{derrida1993exact,blythe2000exact}.
ASEP is a 1d lattice gas model of particles hopping in a preferred 
direction with hard core exclusion imposed.
The rate for the particle hopping to the left site and the right site 
is $1$ and $q$, respectively.

Let us consider a 1d lattice with $k$ 
sites where each site can be empty or occupied
by a particle. Then, a configuration of the system is 
specified by a $k$-tuple 
$C = (s_1 , s_2 , \cdots, s_k )$ where $s_i=+$ if
the site $i$ is occupied and $s_i=-$ if the site $i$ is empty.
The probability $P(C)$ for the configuration
$C$ is determined by a Markov process
\begin{equation}
\begin{aligned}
 \frac{d}{dt}|P\ket=H|P\ket
\end{aligned} 
\end{equation}
where $|P\ket$ is given by
\begin{equation}
\begin{aligned}
 |P\ket=\sum_CP(C)|C\ket=\sum_{s_i=\pm}P(s_1,\cdots,s_k)|s_1,\cdots,s_k\ket.
\end{aligned} 
\label{eq:P-exp}
\end{equation}
It turns out that the Markov matrix $H$ of ASEP is given by the Hamiltonian
of the open XXZ spin chain
\cite{Crampe:2014aoa}, where we identify $s_i=\pm$ as the 
spin up and spin down at site $i$.
As shown in \cite{derrida1993exact,blythe2000exact}, the steady state $H|P\ket=0$ of ASEP
is written as
\begin{equation}
\begin{aligned}
 |P\ket=\frac{1}{Z}|\Psi\ket,
\end{aligned} 
\label{eq:P-Psi}
\end{equation} 
where $|\Psi\ket$ is a matrix product state (MPS)
\begin{equation}
\begin{aligned}
 |\Psi\ket=\sum_{s_i=\pm}
\bra W|\h{a}_{s_1}\cdots \h{a}_{s_k}|V\ket
|s_1,\cdots,s_k\ket,
\end{aligned} 
\label{eq:MPS}
\end{equation} 
and $Z$ in \eqref{eq:P-Psi} is a normalization factor
which is introduced to make the total probability equal to one
\begin{equation}
\begin{aligned}
 \sum_{s_i=\pm}P(s_1,\cdots, s_k)=1.
\end{aligned} 
\label{eq:total-prob}
\end{equation}
$\h{a}_\pm$ in \eqref{eq:MPS} is related to 
the $q$-deformed oscillator $a_\pm$
in \eqref{eq:q-osci} by
\begin{equation}
\begin{aligned}
 \h{a}_\pm =a_\pm+\hf E_0,
\end{aligned} 
\label{eq:hat-apm}
\end{equation}
 and $\bra W|$ and $|V\ket$ in \eqref{eq:MPS}
are coherent states of $a_\pm$
\begin{equation}
\begin{aligned}
\bra W|a_{+}=w\bra W|,\qquad
a_{-}|V\ket=v|V\ket.
\end{aligned} 
\label{eq:coherent}
\end{equation}
Note that the operators $D$ and $E$ 
in \cite{derrida1993exact} are related to $\h{a}_\pm$
in \eqref{eq:hat-apm} as
\begin{equation}
\begin{aligned}
 D=\frac{\h{a}_{-}}{\rt{1-q}},\quad E=\frac{\h{a}_{+}}{\rt{1-q}}.
\end{aligned} 
\end{equation}
From the commutation relation of $q$-oscillators \eqref{eq:q-osci},
one can easily show that $D$ and $E$ obey the so-called DEHP
algebra \cite{derrida1993exact} 
\begin{equation}
\begin{aligned}
 DE-qED=D+E.
\end{aligned} 
\end{equation}
When $w,v\ne0$, the boundary states $\bra W|$ and $|V\ket$ 
in \eqref{eq:coherent} represent the end-of-the-world brane of DSSYK
\cite{Okuyama:2023byh}.
In this paper, we set $w=v=0$ for simplicity. Then 
the state $|\Psi\ket$ in \eqref{eq:MPS} becomes
\begin{equation}
\begin{aligned}
  |\Psi\ket=\sum_{s_i=\pm}
\bra 0|\h{a}_{s_1}\cdots \h{a}_{s_k}|0\ket
|s_1,\cdots,s_k\ket.
\end{aligned} 
\label{eq:MPS2}
\end{equation}

Our key observation is that the normalization condition \eqref{eq:total-prob}
is written as
\begin{equation}
\begin{aligned}
 \bra \Om|P\ket=1,
\end{aligned} 
\label{eq:Om-P}
\end{equation} 
where $\bra\Om|$ is a product state
\begin{equation}
\begin{aligned}
 \bra\Om|=\bra \mathbbm{1}|^{\otimes k}=
\overbrace{\bra \mathbbm{1}|\otimes \bra \mathbbm{1}|\otimes\cdots 
\otimes \bra \mathbbm{1}|}^{k},\qquad
\bra \mathbbm{1}|=\bra +|+\bra -|.
\end{aligned} 
\end{equation}
From \eqref{eq:Om-P}, it follows that $Z$ 
in \eqref{eq:P-Psi} is written as
the overlap of $\bra\Om|$ and $|\Psi\ket$
\begin{equation}
\begin{aligned}
 Z=\bra\Om|\Psi\ket.
\end{aligned} 
\label{eq:ZASEP-strange}
\end{equation}
Using \eqref{eq:MPS2},
we can see that $Z$ is written as a moment of 
the transfer matrix $T=a_{+}+a_{-}$
\eqref{eq:T} of DSSYK 
\begin{equation}
\begin{aligned}
 Z=\bra 0|(\h{a}_{+}+\h{a}_{-})^k|0\ket=\bra 0|(T+E_0)^k|0\ket,
\end{aligned} 
\label{eq:Z-ASEP}
\end{equation}
up to a constant shift $T\to T+E_0$.
This shift is necessary for the probability interpretation of $|P\ket$:
the spectrum of $T+E_0$ is positive definite 
since the eigenvalues of $T$ in \eqref{eq:Eth}
are distributed along the cut $E(\th)\in [-E_0,E_0]$.
The thermal partition function of DSSYK in \eqref{eq:Z-disk}
is obtained by formally 
considering the generating function of $Z$ in \eqref{eq:Z-ASEP}
with respect to the length $k$ of the spin chain
\begin{equation}
\begin{aligned}
 \sum_{k=0}^\infty\frac{(-\bt)^k}{k!}\bra 0|(T+E_0)^k|0\ket
=e^{-\bt E_0}\bra 0|e^{-\bt T}|0\ket.
\end{aligned} 
\end{equation}
This agrees with $Z(\bt)$ in \eqref{eq:Z-disk} up to an overall factor
$e^{-\bt E_0}$.
By the abuse of terminology, we will call $Z$ in \eqref{eq:Z-ASEP}
as the partition function.
Interestingly, $Z$ in \eqref{eq:ZASEP-strange}
takes the same form as the strange correlator appearing in the 
Turaev-Viro/string-net correspondence \cite{You2013WaveFA,Vanhove:2018wlb}
as we explain in the next section.
This strongly suggests that the ASEP/DSSYK duality
can be thought of as
a lower dimensional analogue of the Turaev-Viro/string-net correspondence,
as summarized in Table \ref{tab:correspondence}. 

\section{Turaev-Viro/string-net correspondence}\label{sec:TV}
In this section, we will review the 
Turaev-Viro/string-net correspondence.
This section is just a review of known results and 
contains no new result. However, we will clarify more details about
the analogy between 
the ASEP/DSSYK duality and the Turaev-Viro/string-net correspondence
in Table \ref{tab:correspondence}. 

The 
Turaev-Viro/string-net correspondence 
was first observed in \cite{Kadar:2009fs,Koenig:2010uua}
and later proved mathematically in \cite{kirillov2011stringnetmo}.
This correspondence was further generalized in \cite{Aasen:2020jwb,Lootens:2020mso}
to include defects and domain walls.
The state sum model of
Turaev-Viro \cite{turaev1992state}
and Barrett-Westbury \cite{Barrett:1993ab} can be defined for 
any fusion category $\cC$.
The simple objects $a,b,c$ in the fusion category $\cC$ satisfy the fusion rule
\begin{equation}
\begin{aligned}
 a\otimes b=\bigoplus_{c}N_{ab}^c\, c
\end{aligned} 
\end{equation} 
with some non-negative integer $N_{ab}^c$.
Physically, each  simple object of the category $\cC$ is identified with a
different species of anyons.
For instance, the category for the Ising model has three elements
$\{\mathbbm{1},\psi,\si\}$ and the non-trivial fusion rules are given by
\begin{equation}
\begin{aligned}
 \psi\otimes\psi=\mathbbm{1},\quad
\si\otimes\si=\mathbbm{1}\oplus\psi,\quad
\psi\otimes\si=\si.
\end{aligned} 
\end{equation}
We also need the so-called $F$-symbol 
$[F^{abc}_d]_{x,y}$ for the 
$F$-move of fusion diagram
\begin{equation}
\begin{aligned}
 \begin{tikzpicture}[scale=1]
\draw[thick] (0,0)--(0.6,0);
\draw[thick] (0,0)--(-0.5,0.6);
\draw[thick] (0,0)--(-0.5,-0.6);
\draw[thick] (0.6,0)--(1.1,0.6);
\draw[thick] (0.6,0)--(1.1,-0.6);
\draw (0.3,0) node [below]{$x$};
\draw (-0.5,-0.7) node [below]{$a$};
\draw (-0.5,0.6) node [above]{$b$};
\draw (1.1,-0.6) node [below]{$d$};
\draw (1.1,0.6) node [above]{$c$};
\draw (1.3,0) node [right]{$\displaystyle =~\sum_y \bigl[F^{abc}_d\bigr]_{x,y}$};
\draw[thick] (4,0.6)--(4.3,0.25);
\draw[thick] (4.6,0.6)--(4.3,0.25);
\draw[thick] (4.3,0.25)--(4.3,-0.25);
\draw[thick] (4.6,-0.6)--(4.3,-0.25);
\draw[thick] (4,-0.6)--(4.3,-0.25);
\draw (4,0.6) node [above]{$b$};
\draw (4.6,0.6) node [above]{$c$};
\draw (4,-0.7) node [below]{$a$};
\draw (4.6,-0.6) node [below]{$d$};
\draw (4.3,0) node [right]{$y\quad,$};
\end{tikzpicture}
\end{aligned} 
\label{eq:F}
\end{equation}
and the $F$-symbols should satisfy the pentagon identity
(see e.g. \cite{Aasen:2020jwb,Lootens:2020mso} for details).

Now, let us consider the Turaev-Viro state sum on the 3-manifold $\Si\times I$,
where $\Si$ is a two-dimensional surface and $I=[0,1]$ is the interval.
We draw a trivalent fusion diagram $\cF$ on $\Si$
corresponding to the cell decomposition of $\Si$.
The fusion diagram $\cF$ consists of vertices, edges, and faces (or plaquettes),
and we assign an object $a\in\cC$ to each edge of the diagram $\cF$.  
We can define the Levin-Wen string-net Hamiltonian \cite{Levin:2004mi}
acting on the state $|\cF\ket$ constructed from the fusion diagram
\cite{kirillov2011stringnetmo}
\begin{equation}
\begin{aligned}
 H=\sum_v (1-A_v)+\sum_p (1-B_p),
\end{aligned} 
\label{eq:LW-H}
\end{equation}
where the sums range over the vertices and plaquettes
of $\cF$. In the context of Levin-Wen model, the fusion diagram $\cF$ is called the string-net.
$A_v$ and $B_p$ in \eqref{eq:LW-H} are projectors
\begin{equation}
\begin{aligned}
 A_v^2=A_v,\quad B_p^2=B_p,
\end{aligned} 
\end{equation}
which are defined as follows.
The vertex term $A_v$ is the projector enforcing the
fusion rules
\begin{equation}
\begin{aligned}
 A_v=\sum_{a,b,c\in v}\cob_{abc}|abc\ket\bra abc|,
\end{aligned} 
\end{equation}
where $\cob_{abc}$ specifies allowed ($\cob_{abc}=1$) and 
forbidden ($\cob_{abc}=0$)
configurations of edges incident to a vertex $v$.
$B_p$ is graphically represented as
\begin{equation}
\begin{aligned}
 \begin{tikzpicture}[scale=1]
\draw (0,0) circle[radius=0.8];
\draw[fill] (0,0) circle[radius=0.05];
\draw (0,-0.05) node [below]{$p$};
\draw (0,0.8) node [above]{$a$};
\draw (-0.83,0) node [left]{$\displaystyle B_p=\frac{1}{D^2}\sum_a d_a$};
\end{tikzpicture}
\end{aligned} 
\label{eq:Bp}
\end{equation}
where $d_a$ is the quantum dimension of the object $a\in\cC$ and
$D=\rt{\sum_a d_a^2}$ is the total dimension.
In \eqref{eq:Bp}, we have replaced the plaquette $p$
of the original lattice $\cF$ by a vertex of the dual lattice $\cF^{\vee}$.
The string-net state $|\Psi\ket$ is characterized as a ground state of the 
Levin-Wen Hamiltonian in \eqref{eq:LW-H}
\begin{equation}
\begin{aligned}
 H|\Psi\ket=0.
\end{aligned} 
\end{equation}
As shown in \cite{Gu2008TensorproductRF,Buerschaper:2008eyf},
the string-net state $|\Psi\ket$ is written as a PEPS
(projected entangled pair state)
\begin{equation}
\begin{aligned}
 |\Psi\ket=\prod_p B_p|\Om\ket,
\end{aligned} 
\label{eq:string-net}
\end{equation}
where $|\Om\ket$ is a product state obtained by assigning the 
unit object $\mathbbm{1}\in\cC$ to all edges of $\cF$
\begin{equation}
\begin{aligned}
 |\Om\ket=\prod_e|\mathbbm{1}\ket_e.
\end{aligned} 
\label{eq:Om-e}
\end{equation}
Note that the Kitaev's toric code
\cite{Kitaev:1997wr} is an example of string-net state based on the category
$\text{Vec}_{\mathbb{Z}_2}$ of $\mathbb{Z}_2$-graded vector spaces.

Finally, the Turaev-Viro state sum on the 3-manifold $\Si\times I$
is written as a strange correlator, i.e. the overlap of 
the string-net state $|\Psi\ket$ in \eqref{eq:string-net}
and the product state $|\Om\ket$ in \eqref{eq:Om-e}
\begin{equation}
\begin{aligned}
 Z_\text{TV}(\Si\times I)=\bra\Om|\Psi\ket.
\end{aligned} 
\label{eq:ZTV-strange}
\end{equation}
Interestingly, this Turaev-Viro state sum
\eqref{eq:ZTV-strange} agrees with the partition function of a
statistical mechanical 2d lattice model on $\Si$ \cite{Aasen:2020jwb}
\begin{equation}
\begin{aligned}
  Z_\text{TV}(\Si\times I)=Z_\text{2d}(\Si),
\end{aligned} 
\label{eq:Z2d-holography}
\end{equation}
where the Boltzmann weight of this lattice model is determined by the $F$-symbol
of the fusion category \eqref{eq:F}.
One can clearly see the similarity between the partition function of
ASEP/DSSYK in \eqref{eq:ZASEP-strange} and
the Turaev-Viro state sum in \eqref{eq:ZTV-strange}.
The precise dictionary is summarized in Table \ref{tab:correspondence}.

\section{Discussion}\label{sec:discussion}
We have seen that there is an interesting similarity between 
the ASEP/DSSYK duality in section \ref{sec:ASEP}
and the Turaev-Viro/string-net correspondence 
in section \ref{sec:TV}, as summarized in Table \ref{tab:correspondence}. 
There are many interesting open questions about this analogy, 
 which deserve further study.
We will discuss some of them below.
Also, based on this analogy, we expect that we can draw some lessons
for the holography of 3d gravity from the knowledge
of how the holography works in the lower dimensional case
of DSSYK.

\paragraph{Gapped bulk $H$ and boundary CFT}
Usually, the spectrum of the Levin-Wen Hamiltonian $H$
is assumed to be gapped, which makes the ground space $H|\Psi\ket=0$
of $H$ topologically protected.
This protected property allows us to use the string-net state in
a fault tolerant quantum computation, known as the  
Turaev-Viro codes \cite{Koenig:2010uua,Bonesteel:2012pkv,Schotte:2020lnz}.
It would be interesting to understand the relation between the Turaev-Viro codes
and the holographic error correcting codes 
\cite{Almheiri:2014lwa,Pastawski:2015qua}, if any.

We should stress that the gapped spectrum of $H$ does not contradict
the AdS/CFT correspondence, where the boundary theory is described by a gapless
CFT. As shown in Figure \ref{fig:sandwich},
$H$ and the transfer matrix $T$ of the boundary theory
are different symmetry generators of the bulk geometry $\Si\times I$:
$H$ generates the translation along $I$ while $T$ generates the translation along $\Si$.
Since $T\ne H$,
it is completely consistent if the transfer matrix $T$ of 
the boundary theory is gapless while the 
spectrum of the bulk $H$ is gapped.

\paragraph{Off-critical holography}
In the AdS/CFT correspondence, we usually assume that
the boundary theory is a critical theory described by a
conformal field theory.
However, the holographic duality is a more broader concept which 
might work even in the off-critical regime.
In fact, the holography of DSSYK is an example of this type of 
off-critical duality.
If we take the double-scaling limit near the spectral edge
$E(\th)=-E_0$ of DSSYK, we find the near $\text{AdS}_2/\text{CFT}_1$
holography between the JT gravity and the Schwarzian theory
\cite{Maldacena:2016upp}, or its
higher genus generalization of the double-scaled matrix model
of Saad-Shenker-Stanford \cite{Saad:2019lba}.
It is argued in \cite{Blommaert:2024ydx,Blommaert:2024whf}
that the holographic dual of DSSYK is 
given by the sine dilaton gravity, and this duality works even 
away from the spectral edge
$E(\th)=-E_0$ and the sine dilaton gravity is dual to DSSYK
including the whole spectrum $-E_0\leq E(\th)\leq E_0$.
In other words, the holography between the sine dilaton gravity
and DSSYK works without taking a critical limit 
(or the double-scaling limit of the matrix model).

This is similar to the situation of the holography between 2d lattice
statistical model and 3d Turaev-Viro state sum in 
\eqref{eq:Z2d-holography}.
This duality \eqref{eq:Z2d-holography} is an exact relation
between $Z_\text{2d}(\Si)$ and $Z_{\text{TV}}(\Si\times I)$
which is valid even away from the critical CFT limit of 
$Z_\text{2d}(\Si)$.
There are many example of 2d lattice models which flow to 
CFT in a certain critical limit.
For instance, the Andrew-Baxter-Forester model
\cite{Andrews:1984af} corresponds to the $(p,p+1)$ unitary minimal
model CFT in the critical limit.
However, $Z_\text{2d}(\Si)$ and $Z_{\text{TV}}(\Si\times I)$
in \eqref{eq:Z2d-holography}
are written in terms of the data of fusion category which
make sense even away from the critical regime.
For instance, the partition function of the Ising model
was studied extensively in \cite{Aasen:2016dop}
from the viewpoint of fusion category without taking the critical 
CFT limit. We expect that the relation $Z=\bra\Om|\Psi\ket$
for the ASEP/DSSYK \eqref{eq:ZASEP-strange} and
the Turaev-Viro/string-net \eqref{eq:ZTV-strange}
can be applied to a more general case of holography even 
in the off-critical regime of the boundary theory.

Since we are not taking the continuum limit of boundary theory, the
bulk geometry has some built-in discreteness.
In the case of DSSYK, the chord number $n$ 
in \eqref{eq:number} corresponds to the 
bulk geodesic length, which becomes quantized in units of $\la$ in 
\eqref{eq:limit} \cite{Berkooz:2018jqr,Lin:2022rbf}.
Thus, $\la$ can be thought of as the Planck length of the 
bulk quantum gravity of DSSYK.
It would be interesting to understand the Planck scale structure
of the Turaev-Viro/string-net
correspondence and see how the continuous bulk geometry appears in the critical
limit of the 2d boundary theory on $\Si$.

\paragraph{Branes and defects}
The stationary state $|\Psi\ket$ of the ASEP in \eqref{eq:MPS} 
is written as a MPS with the boundary conditions
$\bra W|$ and $|V\ket$.
As discussed in \cite{Okuyama:2023byh}, these boundary states
correspond to the end-of-the-world branes in the bulk dual 
of DSSYK. In the case of Vuraev-Viro state sum, we can also introduce boundaries
on the two-dimensional surface $\Si$. Then the corresponding 3d bulk theory is
described by an extended TQFT \cite{balsam2012}.
It would be interesting to understand the branes in the bulk 3d geometry
$\Si\times I$ in this case of Turaev-Viro/string-net correspondence
with boundaries.

We can also introduce defects in the Turaev-Viro/string-net correspondence.
Then the partition function becomes
\begin{equation}
\begin{aligned}
 Z_\text{defect}=\bra\mathcal{D}|\Psi\ket,
\end{aligned} 
\end{equation}
where $|\cD\ket$ is obtained by acting some defect creation operator
$\cD_\varphi$ on the product state $|\Om\ket$ in \eqref{eq:Om-e}.
As discussed in \cite{Aasen:2020jwb}, the 
defect creation operator
$\cD_\varphi$ commutes with the transfer matrix $T$ of the 2d lattice model
\begin{equation}
\begin{aligned}
 [\cD_\varphi,T]=0.
\end{aligned} 
\end{equation}
This is 
similar to the relation \eqref{eq:bilocal-T}
for the bi-local matter operator $\cD_{\bt,\lap}$ in DSSYK.
It would be interesting to better understand the relation between 
the defect creation operator in Turaev-Viro/string-net and the bi-local
matter operator in ASEP/DSSYK.
 
\paragraph{Sum over bulk topologies} In the holography of ASEP/DSSYK and
Turaev-Viro/string-net, the bulk geometries have a fixed topology:
the bulk dual of DSSYK is a disk while the bulk  
geometry of Turaev-Viro/string-net is $\Si\times I$.
On general grounds, we expect that the sum over topologies in the bulk 
arises if we consider some averaging on the boundary theory side.
In fact, in the case of DSSYK one can introduce the so-called ETH matrix model
\cite{Jafferis:2022wez} by replacing the transfer matrix $T$ with a 
random matrix $M$.
Then the moment $\bra0|T^k|0\ket$ of the transfer matrix $T$
is replaced by the average of $\Tr M^k$ in the $N\times N$ random matrix ensemble
\begin{equation}
\begin{aligned}
\bra\Tr M^k\ket =\frac{\int dM \Tr M^k e^{-N\Tr V(M)}}{\int dM e^{-N\Tr V(M)}},
\end{aligned} 
\label{eq:TrM}
\end{equation} 
where the potential is expanded in terms of the Chebyshev polynomials
$\si_b=\frac{2}{b}\Tr T_b(M/E_0)$
\begin{equation}
\begin{aligned}
 \Tr V(M)=\sum_{r=0}^\infty (-1)^{r-1}q^{\hf r(r+1)}(\si_{2r}-\si_{2r+2}).
\end{aligned} 
\end{equation}
Then the large $N$ genus-expansion of $\bra\Tr M^k\ket$ in \eqref{eq:TrM}
corresponds to the sum over topologies on the bulk side
\begin{equation}
\begin{aligned}
 \bra\Tr M^k\ket =\sum_{g=0}^\infty N^{1-2g}\bra\Tr M^k\ket_g,
\end{aligned} 
\end{equation}
where the genus-zero term $\bra\Tr M^k\ket_{g=0}$
reproduces the moment $\bra0|T^k|0\ket$ of DSSYK.

We expect that a similar story holds for the Turaev-Viro/string-net 
correspondence.
In fact, the $q$-deformed oscillators also appear in the $Q$-operator of 
2d integrable lattice model  
\cite{Bazhanov:1996dr,Bazhanov:1998dq,Bazhanov:2022wdj}.
It turns out that the occupation number of  
the $q$-deformed oscillator is interpreted as the height variable of a 2d lattice
model, known as the
height model, via the shadow world construction of Turaev-Viro state sum
\cite{turaev1992shadow,Aasen:2020jwb}.
From the analogy with DSSYK,
it is tempting to speculate that the sum over geometries in 3d Turaev-Viro 
state sum naturally arises by replacing the $q$-oscillators in the transfer matrix
of 2d lattice model with random matrices.
This can be thought of as replacing the PEPS of string-net state by a random
tensor network \cite{Qi:2018shh}.
It would be interesting to construct this random tensor network
explicitly.

\paragraph{Comparison with other approaches}
We should stress that our approach is different from the 3d gravity based on the
Virasoro TQFT 
\cite{Collier:2023fwi,Collier:2024mgv,Hartman:2025cyj,Hartman:2025ula}.
In that case, the corresponding 2d boundary theory is 
some ensemble average of CFTs.
See also \cite{Belin:2020hea,Chandra:2022bqq,Belin:2023efa,Jafferis:2025vyp,Jafferis:2025jle,Dymarsky:2024frx}
for the averaging of CFTs and its relation to 3d gravity.
The sum over geometries in 3d gravity is discussed in 
\cite{Yan:2025usw,deBoer:2025rct,Stanford:2025llj,Dymarsky:2026lnf,Belin:2026pko}
in this setting.
Our Turaev-Viro/string-net correspondence 
\eqref{eq:Z2d-holography} is an exact duality without taking 
the ensemble average. As we mentioned above, we can
introduce averaging by replacing the string-net state by a 
random tensor network. However, this random tensor network is
defined at the level of the lattice model and hence it is different from the
averaging of CFTs. Also, our approach is different from the one in
\cite{Bao:2024ixc}; we are not considering the RG flow to CFT since our 
holographic duality works in the off-critical regime as we emphasized above.

\acknowledgments
The author would like to thank Mengyang Zhang for correspondence.
This work was supported
in part by JSPS KAKENHI 25K07300. 
A preliminary result of this paper was presented in the workshop ``Double Scaled Sachdev-Ye-Kitaev Model: From Gravity to Many-Body Quantum Chaos'' at Simons Center for Geometry and Physics on May 13, 2026.

\bibliography{cite}
\bibliographystyle{utphys}

\end{document}